# LIGHT BY LIGHT SCATTERING AT A LOW ENERGY $e^+e^-$ COLLIDER
## or What to do while waiting for that rare kaon decay event

M.R. Pennington

Institute for Particle Physics Phenomenology, University of Durham, Durham DH1 3LE, U.K.

*Abstract*

Two photon processes are a wonderful by-product of an $e^+e^-$ collider at any energy. Two photon reactions probe the very structure of matter. A machine operating below 2 GeV has the potential to reveal the Higgs sector of the strong interaction — the scalars with vacuum quantum numbers that reflect dynamical mass generation from the breakdown of chiral symmetry. One of the many $e^+e^-$ machines around the globe must have a dedicated $\gamma\gamma$ facility.

## INTRODUCTION

When two charged leptons approach each other and they each feel each other's electromagnetic field, there is a probability that they each radiate a photon and these collide to produce hadrons, Fig. 1. Because of the dynamics of the photon propagator, the most likely situation is that these photons are as close to mass-shell as possible. Since we know the nature of the lepton-photon vertices, we can factor these off and as on the right hand side of Fig. 1 learn about real two photon interactions.

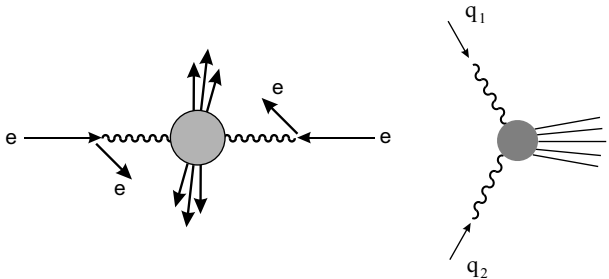

Figure 1: Charged leptons colliding produce two photon reactions, where the photon momenta $q_1$, $q_2$ are very nearly on-shell.

Why do we want to do this? Because we want to learn about the Higgs sector of the fundamental interactions. Once the Higgs of electroweak symmetry breaking has been discovered at the LHC, the next step is to study its properties in detail. Consequently, at the Next Linear Collider there are ingenious plans to generate two photon interactions to probe the nature of this Higgs. At lower energy $e^+e^-$ colliders we can study the Higgs sector of the strong interaction. There the scalar or scalars are responsible for the breaking of chiral symmetry, for the generation of constituent quark masses and consequently the masses of all light hadrons. As we shall recall, there appears not one such Higgs, but a whole series of scalars with vacuum quantum numbers that overlap with each other in a complex way. Of course, this may well be a guide to what the Higgs sector of electroweak symmetry breaking may be like too.

The reach of two photon physics clearly depends on the energy of the $e^+e^-$ collider and the precision possible depends on the luminosity available. The probability of two photon collisions increases with the square of the logarithm of the energy. A machine running at just 750 MeV to 1 GeV per beam allows a number of key channels to be explored: individual hadron states like the $\pi^0$, $\eta$ and $\eta'$, and crucial two body channels, $\pi\pi$, $\pi^0\eta$ and $K\overline{K}$. What do we learn from these?

## INDIVIDUAL MESONS

We start by modelling the formation of individual quark model states in two photon collisions by a simple quark exchange picture, Fig. 2. Then the rate depends on the square of the average squared charge of the constituent quarks times the probability that quarks are created. In the non-relativistic quark model this probability is proportional to the square of the wavefunction at the origin. Since the $\pi^0$, $\eta$ and $\eta'$ are all made of the same quarks, the mean charges squared are of the same magnitude and so we might expect their rate of formation to be similar. Yet experiment tells us

$$\Gamma(\pi^0 \to \gamma\gamma) : \Gamma(\eta \to \gamma\gamma) : \Gamma(\eta' \to \gamma\gamma) \simeq 1 : 60 : 500 .$$

To understand this, it is not greater experimental precision that we need, but rather genuine non-perturbative calculations of confined quarks that explain, rather than just model [1], this strong dependence ($\sim M_{PS}^3$) on the pseudoscalar ($PS$) mass. To this end we need to put together studies of the quark propagator with calculations of how a quark and antiquark build bound states within QCD [2].

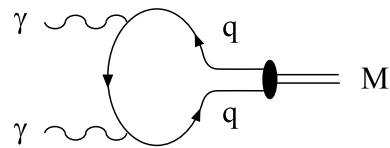

Figure 2: Creation of a neutral meson composed of a $q\overline{q}$ pair by two photons. As a Feynman graph, a contribution interchanging the two photons must be added to respect gauge invariance.

# EXCLUSIVE REACTIONS

It is in two body channels that the remarkable power of two photon reactions to probe the structure of matter comes into its own. In Fig. 3 is shown the measured $\gamma\gamma \to \pi\pi$ cross-section [3, 4, 5, 6, 7] in both the charged and neutral pion case with a c.m. energy from threshold to 2.5 GeV.

Let us look at the dynamics that underlies these cross-sections. Think about the process from the crossed-channel perspective. At low energies, when the photon has long wavelength, it sees the whole hadron and couples to its electric charge. Consequently the charged pion cross-section is large, while that for neutral pions is small. However, as the energy goes up, the wavelength of the photon shortens and it recognises that the pions whether charged or neutral are made of the same constituent quarks and can cause these to resonate. As a result dominating the cross-sections of Fig. 3 is the well-known $q\bar{q}$ resonance, the $f_2(1270)$ — having spin-2 it is readily created by two spin-1 photons. As the energy increases further, the photon has still shorter wavelength and instead of seeing *dressed* or constituent quarks, it couples to almost *bare* or current quarks. As shown first by Brodsky and Lepage [8], these interactions can be calculated perturbatively in QCD. Predictions are in reasonable agreement with experiment already at 2 GeV, as seen in the inset in Fig. 3.

Within just a short energy range the photon reveals dynamics on three different scales, Fig. 3. At the lowest energies the photon couples to the whole hadron. There pion interactions are governed by chiral dynamics and this is embodied in Chiral Perturbation Theory ($\chi$PT). The process $\gamma\gamma \to \pi^0\pi^0$ has no Born contribution, so the lowest order contributions are one loop graphs, the sum of which is finite. This gives the prediction shown in Fig. 4(a), which rises almost linearly with energy [9] and agrees with the only available data from Crystal Ball [5] at just a couple of energies. Since this prediction was advertised as a *gold-plated test of $\chi PT$* [10], the conclusion by some at the time of the DAΦNE-I proposal [11] was that the data must be wrong. However, one can calculate the cross-section non-perturbatively using dispersion relations. This is possible because $\gamma\gamma$ can go to $\pi^+\pi^-$, a process dominated by its Born term at low energies, and then the $\pi^+\pi^-$ can scatter and go to $\pi^0\pi^0$ through final state interactions that are calculable. This cross-section was computed by David Morgan and myself [12, 13] and found to be in agreement with experiment, Fig. 4(b). Since "$\chi$PT never misses" to quote Meißner [14], the problem must be with lowest order of perturbation theory.

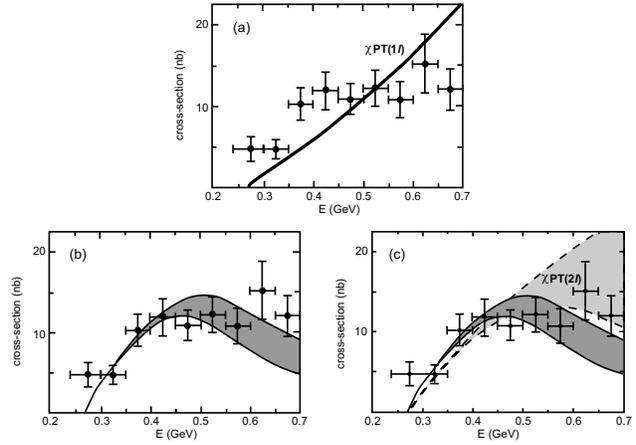

Figure 4: Integrated cross-section for $\gamma\gamma \to \pi^0\pi^0$ as a function of the $\pi\pi$ invariant mass $E$. The data are from Crystal Ball [5] scaled to the full angular range by a factor of 1.25. (a) shows the prediction of Chiral Perturbation Theory at one loop order, labelled $\chi PT(1\ell)$ [9]. (b) shows the dispersive prediction [12, 13]. The shaded band reflects the uncertainties in current experimental knowledge of both $\pi\pi$ scattering and vector exchanges. (c) shows the dispersive band of (b) together with the prediction of Chiral Perturbation Theory at two loops, $\chi PT(2\ell)$ [15], with its shaded band reflecting the uncertainty in the determination of the higher order parameters in the chiral Lagrangian.

One loop $\chi$PT for $\gamma\gamma \to \pi\pi$ involves just tree level $\pi\pi$ interactions and one can easily see that this alone does not accurately reproduce the experimental $I = 0, 2$ $S$-wave phase-shifts [13] that are included in the dispersive calculation. By going to two loop $\chi$PT, Bellucci, Gasser and Sainio [15] found far better agreement with the Crystal Ball data, Fig. 4(c), within the uncertainties in the higher order constants. This diffused the need to re-measure this cross-section, at least below 500 MeV. However, above the near threshold region neither dispersion relations nor higher order $\chi$PT can be reliably computed and we need data. This is not surprising, since only experiment can determine the two photon coupling of resonances, which are a key guide to their constitution.

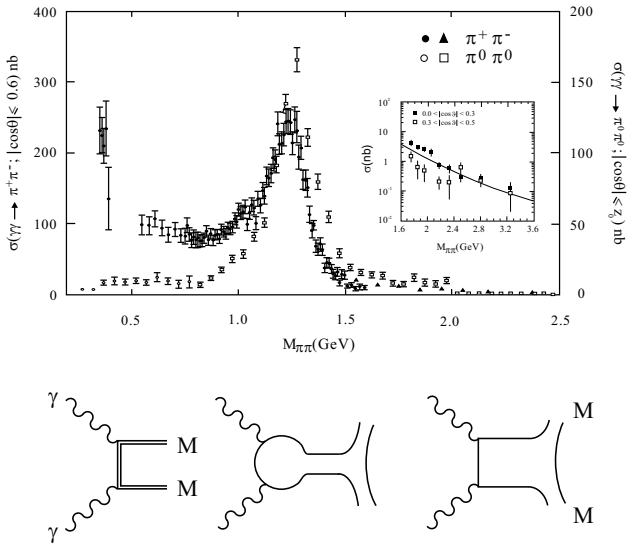

Figure 3: Cross-section for $\gamma\gamma \to \pi\pi$ integrated over $|\cos\theta| \leq z_0$. Solid circles are for $\pi^+\pi^-$ from Mark II [3] with $z_0 = 0.6$. Solid triangles are the sum of $\pi^+\pi^-$ and $K^+K^-$ from CLEO [4]. The open circles and squares are $\pi^0\pi^0$ from Crystal Ball with respectively $z_0 = 0.8$ [5] and (above 0.825 GeV) $z_0 = 0.7$ [6]. The inset displays the higher energy data on $\pi^+\pi^-$ (and $K^+K^-$) from Mark II [7] for two ranges of $\cos\theta$, together with a theoretical prediction from [8]. Below are graphs representing the dominant dynamics in each kinematic region, as described in the text.

The resonances with scalar quantum numbers to be found up to 1.8 GeV in the $\pi\pi$ channel are shown in Fig. 5. They are the $f_0(400-1200)$ (or $\sigma$), $f_0(980)$, $f_0(1370)$, $f_0(1500)$ and $f_0(1710)$. What the couplings of these scalars to $\gamma\gamma$ are will help to unravel which scalar plays what role in the breaking of chiral symmetry, which is a glueball, which is largely $qq\overline{qq}$ and which $q\overline{q}$, and how these reflect the properties of the QCD vacuum.

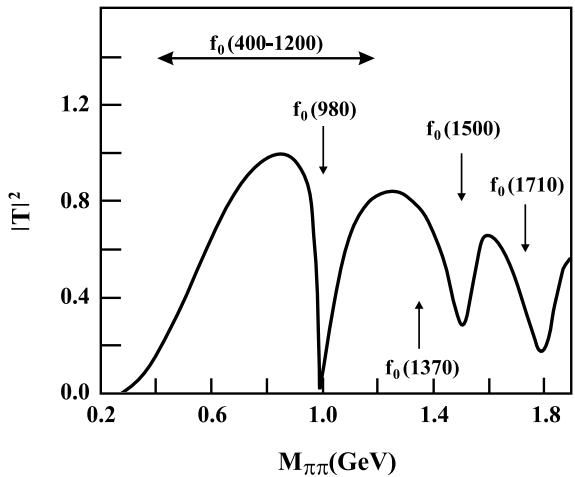

Figure 5: The modulus squared of the $\pi\pi \to \pi\pi$ $I = J = 0$ partial wave amplitude, $T$, indicating the positions of the putative states.

To deduce the radiative widths of resonances, and so gain a handle on their composition, requires a clean separation of data into spin components. With existing data, Fig. 3, the coupling of the $f_2(1270)$ is most constrained, since it dominates the cross-section. However for the scalars that lie underneath this large tensor signal, the uncertainties are large. Without amplitude separation, only with imagination can one see the $f_0(980)$ is there in Fig. 3. If one had data on interactions with polarized photons and complete angular coverage of the hadron final state, a true amplitude analysis would be relatively straightforward. However, we have no polarization information and coverage of typically 60%, and at best 80%, of the angular range in $\cos\theta$ (Fig. 3). It is only for the $\pi\pi$ final state that one can find sufficient other constraints to make a partial wave separation possible. $\gamma\gamma \to \pi^+\pi^-$ is dominated by the Born term at low energies (Fig. 1), as we already commented modified by calculable final state interactions. Moreover unitarity relates the $\gamma\gamma$ process to other hadronic reactions, on which we may have precise experimental information. Below 1.4 GeV or so, before multipion channels start to become important, just the $\pi\pi$ and $K\overline{K}$ intermediate states are all that are needed to impose unitarity. One can thus make up for the inadequacies of the two photon information by incorporating hadronic scattering data into the codes. Even then such an analysis is only possible if data on both $\pi^+\pi^-$ and $\pi^0\pi^0$ final states are included simultaneously. Following earlier studies with David Morgan [16], Elena Boglione and I [17] have completed such an Amplitude Analysis. This reveals two solutions differentiated by whether the $f_0(980)$ appears as a peak or as a dip, Fig. 6. For each of these solutions, one can deduce the radiative widths of the $f_2(1270)$, $f_0(980)$ and $f_0(400-1200)$. These are listed in Table 1.

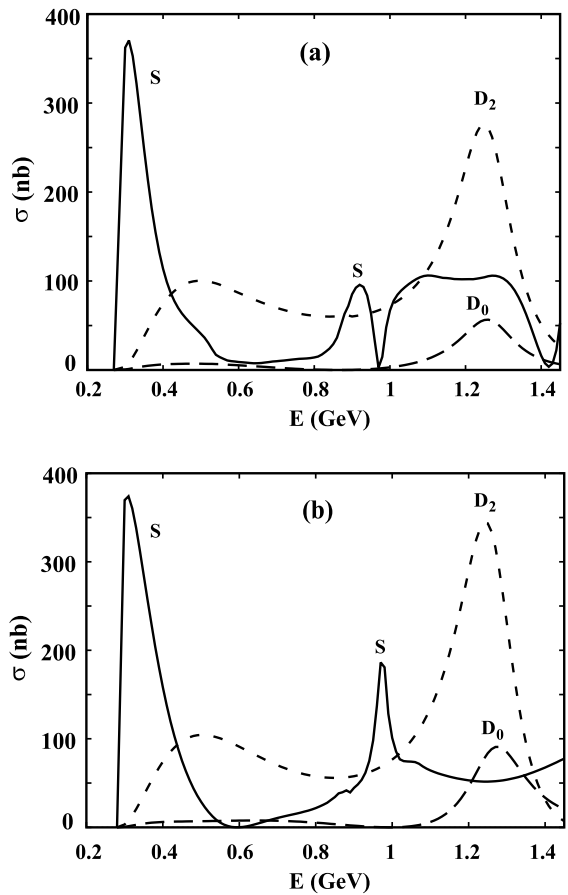

Figure 6: Contributions of the individual partial waves, labelled as $J_\lambda$ (where $J$ = spin and $\lambda$ = helicity) to the integrated $I = 0$ $\gamma\gamma \to \pi\pi$ cross-section describing the data of Fig. 3, from the Amplitude Analysis of [17]. (a) is for the so called *dip* solution, (b) for the *peak* solution. The radiative widths of the corresponding resonances are given in Table 1. The peak close to threshold is due to the $\pi$-exchange Born contribution.

In each solution there is a sizeable $I = 0$ $S$–wave signal through the $f_2$ region, required by the difference in shape of the 1270 MeV peak in the $\pi^+\pi^-$ and $\pi^0\pi^0$ modes, seen in Fig. 3. This $S$–wave signal (Fig. 6) is attributed to the $f_0(400 - 1200)$ with its radiative width given in Table 1.

|  | $\Gamma(\gamma\gamma)$ keV | | |
| --- | --- | --- | --- |
| solution | $f_2(1270)$ | $f_0(980)$ | $f_0(400/1200)$ |
| dip | 2.64 | 0.32 | 4.7 |
| peak | 3.04 | 0.13 – 0.36 | 3.0 |

Table 1: Radiative widths of states contributing to $\gamma\gamma \to \pi\pi$ below 1.4 GeV for the two solutions (*dip* and *peak*) found in the Amplitude Analysis of [17].

| keV | $n\bar{n}$ | $s\bar{s}$ | $K\bar{K}$ |
|---|---|---|---|
| $\Gamma(0^{++} \to \gamma\gamma)$ | 4.5 | 0.4 | 0.6 |

Table 2: How the predicted $\gamma\gamma$ width of scalars in the 1–1.3 GeV region depends on their composition [18, 16].

For the scalars, these widths should be compared with the predictions in Table 2 for different compositions. One sees that for the $f_0(400 - 1200)$, the $\gamma\gamma$ width is consistent with that for a non-strange $q\bar{q}$ state, i.e. $n\bar{n} \equiv (u\bar{u}+d\bar{d})/\sqrt{2}$, while the two photon width of the $f_0(980)$ is compatible with Barnes' calculations [18] for either an $s\bar{s}$ or $K\bar{K}$–molecular composition. In reality the $f_0(980)$ is likely to have a complicated Fock space, in which both $s\bar{s}$ and $K\bar{K}$ components feature strongly [19]. How to calculate the $\gamma\gamma$ coupling of such complexes is as yet unknown [20]. Nevertheless, precision two photon information, differential as well as integrated cross-sections, can distinguish between these solutions. CLEO took such data some time ago, but unfortunately these have never been finalised.

What about the higher mass scalars such as the claimed glueball candidates $f_0(1500)$ and $f_0(1710)$ [21, 22, 20] shown in Fig. 5? Only by extending Amplitude Analyses up to higher masses can one be certain of separating an $S$–wave signal out from under the larger $D$–wave components and so deduce meaningful results for how small the radiative widths of these states might be. Even if they are glueball candidates, they are unlikely to be pure glue, but inevitably mix with the nearby $q\bar{q}$ multiplet(s). As soon as this happens, these states cease to have tiny radiative widths, but have those typical of $q\bar{q}$ scalars, cf. Table 2.

The challenge for the future is for one of the many $e^+e^-$ colliders round the globe to have a dedicated two photon team committed to delivering accurate measurements, from which the two photon widths of all the low mass scalars can be deduced. This requires the study of all accessible final states, $\pi^0\pi^0$, $\pi^+\pi^-$, $K^+K^-$, $K^0\bar{K}^0$, $4\pi$ (as well as $\pi^0\eta$ to understand the related $I = 1$ sector) with as large an angular coverage as possible. Combining this information with reliable predictions from non-perturbative QCD of the two photon couplings, we can then expect to determine the composition of these key hadrons. Only then will we understand the nature of the light scalar mesons, a nature and composition that is intimately tied to the structure of the QCD vacuum — both $q\bar{q}$ and glue. So while at an upgraded $e^+e^-$ collider, like the $\phi$-factory DA$\Phi$NE, one searches the copious kaon events to find decays in modes with miniscule branching fractions to learn what lies beyond the Standard Model, the prospect exists to use the tedious waiting time to expose the structure of the vacuum of the strong interaction that shapes the femto-universe and keeps quarks and gluons confined — a task for which a two photon capability is essential.

## ACKNOWLEDGMENTS

I am very grateful to Rinaldo Baldini and Gino Isidori for including two photon physics as a topic at this meeting. I acknowledge the partial support of the EU-RTN Programme, Contract No. HPRN-CT-2002-00311, "EURIDICE".

## REFERENCES

[1] C. Hayne and N. Isgur, Phys. Rev. **D25** (1982) 1944.

[2] M.R. Pennington, "Exclusive channels in $\gamma\gamma$ reactions: light at the end of the tunnel?", Proc. *Int. Conf. on the Structure and Interactions of the Photon (Photon '99)*, Freiburg, May 1999, Nucl. Phys. Proc. Suppl. **82** (2000) 291.

[3] J. Boyer *et al.* (Mark II), Phys. Rev. **D42** (1990) 1350.

[4] J. Dominick *et al.* (CLEO), Phys. Rev. **D50** (1994) 3027.

[5] H. Marsiske *et al.* (Crystal Ball), Phys. Rev. **D41** (1990) 3324.

[6] J.K. Bienlein (Crystal Ball), Proc. *IXth Int. Workshop on Photon-Photon Collisions* (San Diego 1992) eds. D. Caldwell and H.P. Paar (World Scientific), p. 241.

[7] J. Boyer *et al.* (Mark II), Phys. Rev. Lett. **56** (1986) 207.

[8] S.J. Brodsky and P.G. Lepage, Phys. Rev. **D22** (1980) 2157.

[9] J. Bijnens and F. Cornet, Nucl. Phys. **B296** (1988) 557; J.F. Donoghue, B.R. Holstein and Y.C. Lin, Phys. Rev. **D37** (1988) 2423.

[10] L. Maiani, in a lecture summarised in pp. 719-732 of Ref. 11.

[11] Proc. *Workshop on Physics and Detectors for DA$\Phi$NE*, ed. G. Pancheri (pub. INFN, Frascati) 1991.

[12] D. Morgan and M.R. Pennington, Phys. Lett. **B 272** (1991) 134.

[13] M.R. Pennington, *DA$\Phi$NE Physics Handbook*, ed. L. Maiani, G. Pancheri and N. Paver (INFN, Frascati, 1992) p. 379-418; *Second DA$\Phi$NE Physics Handbook*, ed. L. Maiani, G. Pancheri and N. Paver (pub. INFN, Frascati, 1995) pp. 169-190.

[14] U. Meißner, title of many seminars given in the 1990's.

[15] S. Bellucci, J. Gasser and M.E. Sainio, Nucl. Phys. **B423** (1994) 80.

[16] D. Morgan and M.R. Pennington, Z. Phys. **C48** (1990) 623.

[17] M. Boglione and M.R. Pennington, Eur. Phys. J. **C9** (1999) 11.

[18] T. Barnes, Phys. Lett. **165B** (1985) 434.

[19] E. van Beveren *et al.*, Z. Phys. **C30** (1986) 615; N.A. Tornqvist, Acta Phys. Pol. **B16** (1985) 503, Z. Phys. **C68** (1995) 647; M. Boglione and M.R. Pennington, Phys. Rev. Lett. **79** (1997) 1998.

[20] M.R. Pennington, Proc. *Workshop on Photon Interactions and the Photon Structure*, Lund, Sept. 1998, eds. G. Jarlskog and T. Sjostrand, Lunds Universiteit, p. 313 [hep-ph/9811276].

[21] C. Amsler *et al.*, (Crystal Barrel), Phys. Lett. **B342** (1995) 433, **B353** (1995) 571.

[22] C. Edwards *et al.* (Crystal Ball@SLAC), Phys. Rev. Lett. **48** (1982) 458.